\documentclass[aps,prl,superscriptaddress,showpacs,twocolumn]{revtex4}
\usepackage{graphicx}
\usepackage{epsfig}
\begin{document}
\title{Spectral properties of Bunimovich mushroom billiards}

\author{B.~Dietz}
\affiliation{Institut f\"{u}r Kernphysik, Technische
Universit\"{a}t Darmstadt, D-64289 Darmstadt, Germany}

\author{T. Friedrich}
\affiliation{Institut f\"{u}r Kernphysik, Technische
Universit\"{a}t Darmstadt, D-64289 Darmstadt, Germany}

\author{M.~Miski-Oglu}
\affiliation{Institut f\"{u}r Kernphysik, Technische
Universit\"{a}t Darmstadt, D-64289 Darmstadt, Germany}

\author{A.~Richter}
\affiliation{Institut f\"{u}r Kernphysik, Technische
Universit\"{a}t Darmstadt, D-64289 Darmstadt, Germany}

\author{F.~Sch\"afer}
\affiliation{Institut f\"{u}r Kernphysik, Technische
Universit\"{a}t Darmstadt, D-64289 Darmstadt, Germany}
\date{\today}
\begin{abstract}
Properties of a quantum mushroom billiard in the form of a
superconducting microwave resonator have been investigated. They
reveal unexpected nonuniversal features such as, e.g., a
supershell effect in the level density and a dip in the
nearest-neighbor spacing distribution. Theoretical predictions for
the quantum properties of mixed systems rely on the sharp
separability of phase space -- an unusual property met by mushroom
billiards. We however find deviations which are ascribed to the
presence of dynamic tunneling.
\end{abstract}


\maketitle Billiards play a central role in the investigation of
systems with regular, chaotic, or mixed dynamics \cite{classbill}.
When quantum chaos established itself as a new field, billiards
were used as a paradigm for both theoretical and experimental
research \cite{quantbill}. The billiard considered in this Rapid
Communication is from the family of mushroom billiards suggested
by Bunimovich \cite{buni_mushroom} as a generalization of the
stadium billiard {\cite{bunistadium}. Compared to conventional
mixed systems they have the particular property that their phase
space is sharply divided into one regular island and the chaotic
sea, whereas usually the islands of regularity are typically
surrounded by a layer of infinitely many islands. In the simplest
case they consist of a semicircular hat with a symmetrically
attached rectangular stem as shown in Fig.~\ref{fig1}(a). All
regular orbits of mushroom billiards are orbits of the semicircle
billiard with a conserved angular momentum which stay in the hat
forever. Orbits of particles with the same angular momentum form a
semicircular caustic. There is one critical caustic, which
rigorously separates the orbits into regular and chaotic ones; its
radius $r_c$ equals half the width of the stem [dotted line in
Fig.~\ref{fig1}(a)]. Particles moving in the hat with a larger or
equal caustic stay there forever, whereas those with a smaller one
eventually enter the stem and therefore are chaotic. To our
knowledge this clear separation of the phase space has been found
before only for classical maps \cite{clearcutsystems}. Due to this
unusual feature of the phase space classical mushroom billiards
are of interest with respect to different aspects
\cite{mushroom_works}. How comes about the separability of the
classical phase space manifest in properties of the corresponding
quantum billiard? This question motivated us to investigate a
quantum mushroom billiard experimentally using the analogy between
quantum and microwave billiards \cite{billexp,stadium}.
\begin{figure}
\epsfig{figure=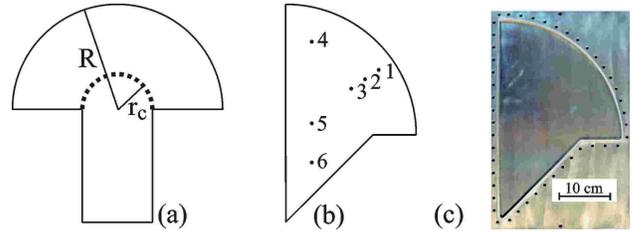,angle=0,height=3cm} \caption{(Color
online) (a) Typical shape of a mushroom billiard with the critical
caustic shown as a dotted line, (b) desymmetrized version
(quarter-circle as hat) with triangular stem, and (c) photograph
of the corresponding experimental microwave resonator. The
positions of the six antennae in the experiment are indicated in
(b).} \label{fig1}
\end{figure}
\par
The geometry of the quantum billiard we investigated is shown in
Fig.~\ref{fig1}(b). To avoid the effects induced by the
superposition of two parity classes, we used a desymmetrized
mushroom billiard -- i.e., one with a quarter-circle for the hat.
Moreover, the stem is chosen triangular instead of rectangular in
order to eliminate bouncing ball orbits as seen, e.g., in the
stadium billiard \cite{stadium}. The stem width is two thirds of
the radius $R$ of the hat, and its inner angle equals $45^\circ$.
We verified with a rigorous analysis \cite{buni_mushroom} that the
classical phase space is still sharply divided into a regular and
a chaotic part. The invariant measure $q_c$ of the chaotic part is
82.9~\% of the phase space volume. The eigenvalues of the quantum
mushroom billiard were measured with a flat, cylindric microwave
cavity of lead plated copper [Fig.~\ref{fig1}(c)] of 5~mm height.
The hat has a radius of $R=0.24$~m. In order to obtain a large and
reliable set of resonance frequencies we performed the
measurements at 4.2~K, where the cavity is superconducting
\cite{stadium,Heine}. Using a vectorial network analyzer (VNA) we
collected complex transmission spectra with six different antennae
up to 22 GHz with a sample rate of 100 kHz. The antenna positions
are distributed over the whole area of the billiard
[Fig.~\ref{fig1}(b)]. Figure~\ref{fig2} shows a part of the
measured transmission spectra for five different antenna
combinations. The spectra have the unusual property that they
exhibit sequences of resonances, which are separated by large gaps
with no resonances (arrows in Fig.~\ref{fig2}). This bunching
effect indicates that there are two well distinguishable frequency
scales. It will be discussed in detail below.
\begin{figure}
\epsfig{figure=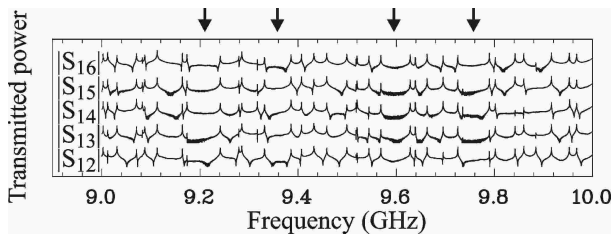,angle=0,width=8cm} \caption{Part of five
transmission spectra between 9 and 10 GHz, measured with the
microwave resonator shown in Fig.~\ref{fig1}(c). The curves
labeled with $|S_{1j}|$, $j=2,...,6$ are obtained with the
antennae $1$ and $j$ whose positions are indicated in
Fig.~\ref{fig1}(b). The spectra have been displaced from each
other along the y axis for illustration and are plotted in a
logarithmic scale. One notices a bunching behavior -- i.e., large
gaps marked by arrows -- between stretches of resonances.}
\label{fig2}
\end{figure}
\par
A set of 938 resonance frequencies was obtained from the spectra
in agreement with the expectation from Weyl's formula
\cite{balteshilf}. On the basis of this large data set we first
considered conventional statistics and compared them to theories
applicable to the quantum spectra of mixed systems, such as, e.g.,
the Berry-Robnik statistics \cite{BerryRobnik}. First we computed
the number of levels $N(f)$ -- i.e.\ the integrated resonance
density $\rho (f)$ -- and determined its smooth part
$N_\mathrm{Weyl}(f)$. The resulting fluctuating part
$N_\mathrm{fluc}(f)=N(f)-N_\mathrm{Weyl}(f)$ is shown in
Fig.~\ref{fig3}. The most striking feature is a beating pattern
caused by the superposition of two oscillations, which in fact
reflects the bunching effect observed in Fig.~\ref{fig2}. In the
stadium billiard \cite{stadium} oscillations are caused by the
so-called bouncing ball orbits. To find the origin of the beating
observed in Fig.~\ref{fig3}, we computed the length spectrum
$\vert\tilde\rho_\mathrm{fluc}(x)\vert$ shown in
Fig.~\ref{fig4}(a), which has peaks at lengths $x$ of periodic
orbits. Here, $\tilde\rho_\mathrm{fluc}(x)$ is the Fourier
transform of the fluctuating part of the resonance density
$\rho_\mathrm{fluc}(k)$ as a function of wave number $k=2\pi f/c$,
where $c$ is the velocity of light. In order to get information on
the nature of the periodic orbits related to the peaks in the
length spectrum, we also computed the latter for the regular
orbits in the hat~\cite{stadium}. For this purpose we first
determined the eigenvalues and eigenfunctions of a quarter-circle
billiard of the same radius as the hat. These are indexed by a
radial quantum number $n$ and an angular momentum one $m$. The
length spectrum in Fig.~\ref{fig4}(b) has been obtained by
considering only eigenvalues with eigenfunctions, which are
localized between the critical caustic [Fig.~\ref{fig1}(a)] and
the circular boundary -- that is, eigenfunctions with a
sufficiently large $m$. By comparison of Fig.~\ref{fig4}(a) with
Fig.~\ref{fig4}(b) we see that all peaks of regular orbits in the
hat of the mushroom billiard are reproduced. The remaining peaks
in the length spectrum are thus associated with chaotic orbits.
Around 0.7 m we observe a pair of closely lying dominant peaks.
They correspond to the shortest regular periodic orbits. After
subtracting their contribution to $N_\mathrm{fluc}(f)$ the beating
vanished. This cause of beatings in circular or spherical
geometries by short dominant orbits of approximately the same
length has already been observed in metal clusters
\cite{metal_clusters} and termed the supershell effect. It has
also been stressed in nuclei \cite{nuclear}, nanowires
\cite{stafford}, and recently in trapped dilute Fermi gases
\cite{aberg}. In mushroom billiards it is induced by regular
orbits coexisting with chaotic ones whose measure is determined by
the depth of the stem.
\begin{figure}
\epsfig{figure=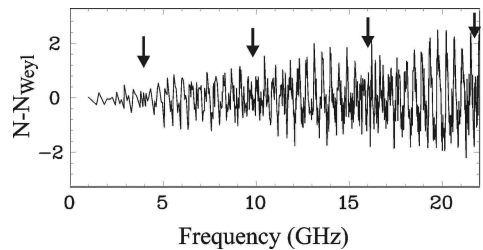,angle=0,height=3.2cm} \caption{Fluctuating
part of the number of resonances below a given frequency $f$. The
curve oscillates and shows a beating behavior with some noise at
the nodes marked by arrows.} \label{fig3}
\end{figure}
\begin{figure}
\epsfig{figure=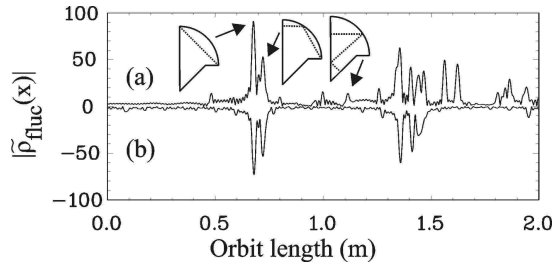,angle=0,height=3.4cm} \caption{(a)
Experimental length spectrum of periodic orbits of the mushroom
billiard. The two peaks of lengths of about 0.7~m are due to two
regular orbits in the hat of the mushroom, and that at length
1.12~m is due to a chaotic one. (b) The computed length spectrum
of a quarter-circle billiard is shown flipped upside down, where
only eigenvalues were considered which correspond to angular
momenta larger than the critical one.} \label{fig4}
\end{figure}
\par
Next we consider the nearest-neighbor spacing distribution (NND).
Figure~\ref{fig5}(a) shows the experimental NND with that for
Poisson statistics and a Gaussian orthogonal ensemble (GOE)
describing regular and chaotic systems, respectively. As to be
expected the experimental NND coincides with neither of them. A
comparison of the NND with the one derived by Berry and Robnik
\cite{BerryRobnik} for mixed systems yields good agreement for
$q_c=$~82.9~\% as given above. Yet, even though the requirements
for its applicability are best met with mushroom billiards, most
surprisingly the agreement is not better than for other mixed
systems. Especially, for small spacings $s$ deviations are visible
in Fig.~\ref{fig5}(a). This can be attributed to dynamic tunneling
between classically separated regions of phase space in the
quantum regime \cite{Heller,tunneling,Reichl}. In addition, we
observe a peculiarity in the shape of the NND. Namely, the
experimental histogram resulting from a fairly small binwidth
exhibits a profound dip at $s=0.7$. This dip is statistically
significant, and it is present also in NNDs derived from different
subsections of the spectrum. When including the contribution of
the pair of dominant short periodic orbits to $N_\mathrm{fluc}(f)$
in the usual spectral unfolding procedure \cite{stadium}, the dip
is no longer present [Fig.~\ref{fig5}(b)]. It should be noted that
this result is contrary to the general belief that only long
periodic orbits control short range spectral properties of a
quantum system.
\begin{figure}
\epsfig{figure=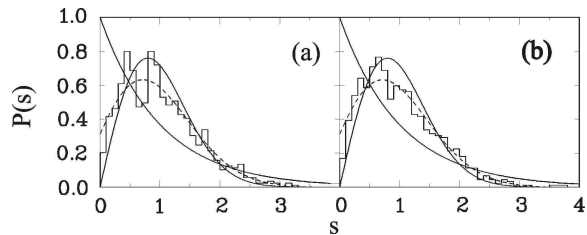,angle=0,height=3cm} \caption{(a)
Nearest-neighbor spacing distribution of the mushroom billiard
together with those for Poisson statistics and the GOE (solid
lines) expected for systems with regular and chaotic dynamics,
respectively. Note the dip at $s=0.7$. It vanishes (b) when the
contribution of the dominant periodic orbits in Fig.~\ref{fig4} to
the resonance density is subtracted. The dashed line in both (a)
and (b) shows the Berry-Robnik prediction for a mixed system with
$q_c=$ 82.9~\% chaos.} \label{fig5}
\end{figure}
\par
While in the semiclassical limit the eigenvalue spectrum of a
mixed system consists of regular eigenvalues with eigenfunctions
localized on the regular islands -- i.e., for the desymmetrized
mushroom billiard those of a quarter-circle billiard with
eigenfunctions localized in the hat -- and of chaotic ones with
eigenfunctions distributed over the whole billiard
\cite{eigenfunktionen}, this must not be true anymore
\cite{Heller,tunneling} in the quantum limit. However, since for
mushroom billiards there is no layer of islands between the
regular and chaotic parts of the phase space, their eigenstates
should be classifiable as regular or chaotic
\cite{Heller,Tomsovic} even in the quantum regime. In order to
investigate this further, we measured the electric field strength
intensity -- i.e., squared eigenfunctions of an enlarged copy of
the mushroom billiard -- for several resonance frequencies using
the perturbing bead method at room temperature \cite{maksim}. An
example for a squared regular eigenfunction is shown in
Fig.~\ref{fig6}(a). It is very similar to that of the
quarter-circle with $n=3$ and $m=44$. A chaotic eigenfunction is
plotted in Fig.~\ref{fig6}(b). However, for such eigenfunctions we
even find traces of regularity in the field pattern in the hat,
and for the regular eigenfunctions the intensity in the stem is
nonvanishing. This indicates that there is a dynamic tunneling
\cite{Reichl}, such that the classification into regular and
chaotic eigenstates is only asymptotically correct. Indeed, we
also observed some rare mixed eigenfunctions whose intensity is
equally distributed over the whole billiard area, while the field
pattern in the hat is very similar to that of a regular
eigenfunction [Fig.~\ref{fig6}(c)]. In Fig.~\ref{fig6}(d) the
averaged intensity distribution of 239 chaotic eigenfunctions is
shown. Its classical counterpart is the probability $P_C$ to find
chaotic orbits at a certain position in the billiard. It is
constant as in completely chaotic systems in that part of the
billiard which is accessible to chaotic orbits only -- i.e., in
the stem and in the hat for $0<r<r_c$, where $r$ is the distance
from the circle center and $r_c$ is the radius of the critical
caustic. Interestingly, for $r_c<r<R$ it decreases as $\arcsin
(r_c/r)$ \cite{buni_mushroom}, which is a consequence of
equipartion in the mixed phase space. The radial profile of $P_C$
is compared to the experimental results in (e) -- revealing the
decrease of intensity in the hat. The abrupt change of the
distribution when crossing the critical caustic is smeared out due
to the finite wavelength. At the boundary of the billiard quantum
effects \cite{arndbaecker} also lead to deviations from the
classical behavior.
\begin{figure}
\epsfig{figure=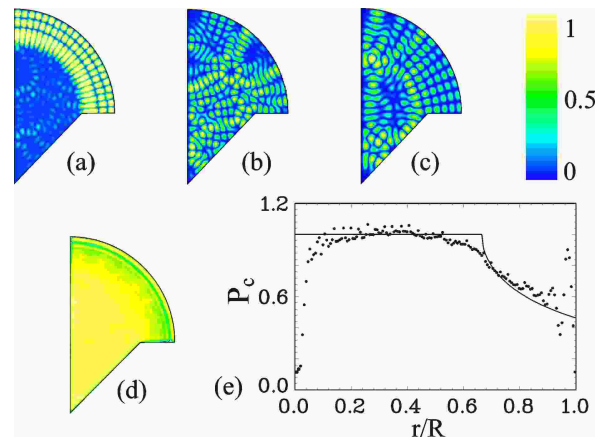,angle=0,width=7.7cm} \caption{(Color
online) Measured intensity distributions. The eigenfunctions are
usually either regular (a) or chaotic (b) and correspond to
eigenfunctions of a quarter-circle billiard in the former case. We
also found some mixed eigenfunctions like the one in (c). In (d)
the averaged distribution of 239 chaotic eigenfunctions is shown.
The radial intensity profile in the hat is plotted in (e) between
the quarter-circle center and the circular boundary (points), and
compared to the classical probability of finding chaotic orbits
(solid line). The decrease beyond the critical caustic at
$r/R=2/3$ is revealed in the measurements, and deviations at the
boundaries and at the knee are due to quantum effects.}
\label{fig6}
\end{figure}
\par
Finally, for a fixed $n$, the eigenvalues of the quarter-circle
become asymptotically equidistant for large frequencies. Thus
regular modes in the superconducting mushroom billiard were
obtained by identifying chains of equidistant eigenvalues in the
measured spectrum and comparing them to the computed ones of the
corresponding circle billiard. We found regular (chaotic) periodic
orbits in the length spectrum of the regular (chaotic) eigenvalues
and, though strongly suppressed, also chaotic (regular) ones
appear. This shows again that there is an interaction via dynamic
tunneling between the regular and chaotic parts of the classical
phase space. The deviations of the measured regular eigenvalues of
the mushroom billiard from the computed eigenvalues of the
quarter-circle billiard shown in Fig.~\ref{fig7} quantify the
dynamic tunneling. Most interestingly they vary unidirectional.
Indeed, for all families of equal radial quantum number $n$ the
deviations are close to zero for large frequencies -- i.e., large
values of $m$ -- where the eigenfunctions of the quarter-circle
billiard are localized close to the circular boundary and
therefore almost not influenced by the stem. When approaching the
frequency corresponding to the critical angular momentum from
above, the deviations increase as then the eigenfunctions are
localized closer and closer to the critical caustic and the field
distribution is increasingly distorted towards the stem as can be
seen in the intensity plot in Fig.~\ref{fig6}(a). This is exactly
the dynamic tunneling. There is no general quantitative
explanation for the measured deviations yet. Our results should,
however, should inspire forthcoming theoretical considerations of
dynamic tunneling.
\begin{figure}
\epsfig{figure=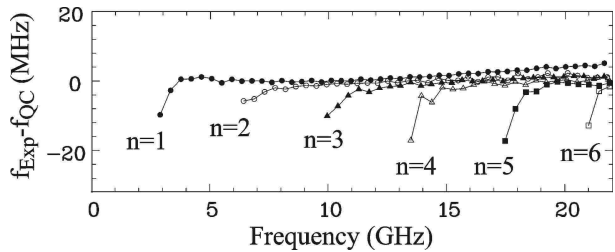,angle=0,width=8cm} \caption{Deviations of
the experimental eigenfrequencies of the mushroom billiard from
those of a quarter-circle billiard versus frequency. The symbols
(points, triangles, squares) refer to regular states. Eigenvalues
with equal radial quantum number $n$ are connected by lines.
For a given $n$ there are almost no deviations for high frequencies
where the eigenfunctions are localized close to the circular
boundary. For smaller frequencies, the influence of the stem leads
to increasing deviations pointing to the importance of dynamic
tunneling.} \label{fig7}
\end{figure}
\par
In summary we have shown that the spectral properties of mushroom
billiards are significantly affected by the shortest regular
orbits in the hat. They cause a supershell structure in the level
density. As compared to other systems exhibiting the supershell
effect, in mushroom billiards the degree of chaoticity can be
tuned by varying the depth of the stem, such that they allow the
study of supershell effects for an arbitrary degree of chaos.
Surprisingly, the shortest (not long) periodic orbits lead to a
substructure in the NND. The eigenstates of the mushroom billiard
may be separated into regular, chaotic, or -- though rare -- mixed
ones. That the latter are rare and the behavior of the averaged
intensity distribution of the chaotic eigenfunctions in the
mushroom hat are manifestations of the separability of the
classical phase space in the spectral properties of the quantum
billiard. Still, dynamic tunneling is present and can be observed,
e.g., in the field distributions, in the spectral properties, and
in the deviations of the regular eigenvalues from those of the
corresponding circle billiard. Since the tunneling barrier is of a
simple structure for mushroom billiards, they are convenient
systems for the study of the dynamic tunneling process.
\begin{acknowledgments} We acknowledge helpful discussions with E.~Bogomolny,
L.~Bunimovich, A.~Heine, T.~Papenbrock, T.~Seligman, and
A.~Wirzba. T.~F. received grants from the Studienstiftung des
Deutschen Volkes. This work has been supported by the DFG within
the SFB 634.
\end{acknowledgments}

\end{document}